\documentclass[prl,twocolumn,showpacs,amsmath,amssymb]{revtex4}

\usepackage{graphicx}       
\usepackage{dcolumn}        

\newcommand{\beq}{\begin{equation}}
\newcommand{\eeq}{\end{equation}}
\newcommand{\beqa}{\begin{eqnarray}}
\newcommand{\eeqa}{\end{eqnarray}}

\newcommand{\kvec}{{\bf k}}

\newcommand{\qvec}{{\bf q}}

\begin{document}
\title{Charge-fluctuation contribution to the Raman response
in superconducting cuprates}

\author{S. Caprara, C. Di Castro, M. Grilli, and D. Suppa}
\affiliation{$^1$Istituto Nazionale di Fisica della Materia,
Unit\`a Roma 1 and SMC Center, and Dipartimento di Fisica\\
Universit\`a di Roma "La Sapienza" piazzale Aldo Moro 5, I-00185 Roma, Italy}

\begin{abstract}
We calculate the Raman response contribution due to
collective modes, finding a strong dependence on the photon polarizations
and on the characteristic wavevectors of the modes. 
We compare our results with recent Raman spectroscopy experiments in  
underdoped cuprates,
${\rm La_{2-x}Sr_xCuO_4}$ and ${\rm (Y_{1.97}Ca_{0.3})Ba_2CuO_{6.05}}$,
where anomalous low-energy peaks are observed,
which soften upon lowering the temperature.  
We show that the specific dependence on doping and on photon polarizations of 
these peaks
is only compatible with charge collective excitations at finite wavelength.
\end{abstract}
\date{\today}
\pacs{74.20.Mn, 78.30.-j, 74.72.-h, 71.45.Lr}
\maketitle

There is an increasing experimental evidence that in the high $T_c$ superconducting
cuprates there are finite-energy excitations, which are distinct from the
usual single-particle excitations characterizing the spectra of standard metals.
In optical conductivity there are many examples of absorption peaks
at finite frequencies, which are quite distinct from the zero-energy Drude peak,
occurring in hole-doped [${\rm La_{2-x}Sr_xCuO_4}$ (LSCO) \cite{lucarelli}, 
${\rm La_{1.6-x}Nd_{0.4}Sr_xCuO_4}$ (LNSCO) \cite{dumm}, ${\rm Bi_2Sr_2CuO_6}$ 
(Bi2201) \cite{lupi}]
as well as in electron-doped materials [${\rm Nd_{2-x}Ce_xCuO_4}$ (NCCO) \cite{singley}].
In all these cases absorption peaks are observed below 200 $cm^{-1}$ which substantially
soften upon decreasing the temperature $T$. It is natural to assign to
these excitations a collective character. Moreover in the case of Bi2201
there is a clear evidence of scaling
properties for the finite-frequency absorption \cite{CDFG}. This suggests
that these excitations are nearly critical fluctuations associated to
some criticality occurring in the cuprates below a critical line $T^*(x)$, ending near
optimal doping at $T=0$. The proposal of
a quantum critical point  in the strongly underdoped \cite{chubukov} or
near optimal doping  \cite{varma,CDG,metzner,localQCP,reviewQCP} is by now acquiring consensus
in the community, but the nature of the ordered phase 
still remains debated. In this regard it is quite important to establish
whether the collective modes (CM) responsible for the finite-frequency absorptions occur
at finite wavevectors and are therefore associated to some form of 
spin and/or charge spatial ordering \cite{chubukov,CDG} or
have infinite wavelength \cite{varma,metzner}.

In order to clarify this issue, in this letter we calculate the Raman
response contribution due to collective charge-order (CO) 
excitations associated to stripe fluctuations and eventually use our results to
interpret recent findings of Raman-scattering experiments \cite{tassini,hacklSNS04}.
Raman scattering is an important tool in the above issue for several reasons. First
of all it is a probe directly accessing the bulk properties in the relevant frequency
range. Most importantly, a suitable choice of the polarization of
the incoming and outgoing photons selects the regions in momentum space
from where excitations are originated \cite{reframan}. As a consequence,
characteristic features arise  in the Raman response 
with distinct doping and polarization dependencies \cite{tassini,hacklSNS04},
which in turn allow to extract valuable informations
about the momentum dependence of the excitations. This is the main issue
of this work. We consider the doping-dependent line $T_{CO}(x)\sim T^*(x)$ \cite{ACDG}.
In the underdoped regime below this line the system would order in the absence
of competing effects. 
Therefore strong CO fluctuations are present in a large portion of the phase diagram
and provide an {\it additional } channel for the
Raman response, besides the usual one obtained from Fermi-liquid quasiparticles (QP).
The role of fluctuations in the response functions was established
by Aslamazov and Larkin for the  paraconductivity in superconductors \cite{AL}.
This theory was extended to the particle-hole channel
for one-dimensional charge-density-wave
systems in Refs.\onlinecite{pattonsham,giapponesi}. In this latter case a partial 
cancellation leads to a less singular paraconductivity as compared to
the particle-particle case. Here we reconsider
this scheme in two dimensions for the Raman response diagrammatically represented
in Figs. 1(a) and 1(b) involving the incommensurate charge CM.
We find that the above cancellation does not generically
occur for the corresponding Raman diagrams, generating important contributions,
which in turn depend on the selected symmetry.
We will argue that only these contributions account
for the {\it strong} symmetry dependence of the Raman response. Other
diagrams with self-energy and vertex corrections are weakly anisotropic
and provide the usual weakly temperature-dependent QP contribution.
\begin{figure}
\vspace{-1 truecm}
\includegraphics[angle=0,scale=0.25]{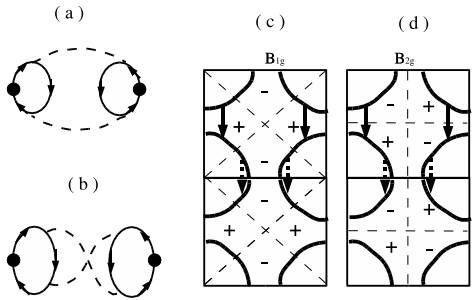}
\vspace{-1 truecm}
\caption{Direct (a) and crossed (b) diagrams for the fluctuation contributions to 
Raman spectra. The dots represent Raman
vertices. Solid lines represent fermionic QP propagators and
dashed lines represent CM propagators. (c) and (d): Hot spots joined by the critical
wavevector ${\bf q}_c=(0,-q_c)$ represented by arrows. Equivalent hot spots
have arrows of the same type (solid or dashed).
The $+$ $(-)$ sign mark the regions where the $B_{1g}$ 
and $B_{2g}$ vertices are positive (negative).
}
\label{fig.1}
%
\end{figure}
Specifically each dashed line appearing in the Raman response in Fig. 1 (a,b)
is the gaussian diffusive CM propagator in Matsubara frequencies 
\beq
D(\qvec, \omega_m)=(|\omega_m|+\Omega_\qvec)^{-1},
\label{diffusive}
\eeq
where $\Omega_\qvec=\nu \vert \qvec -\qvec_c \vert^2 +m(x,T)$ with $\nu$
a constant electronic energy scale (we consider a unit lattice spacing).
$m(x,T)$ is the mass of the CM and encodes the distance
from the critical line $T_{CO}(x)$. This propagator is dominant at 
zero frequency and ${\bf q}={\bf q}_c$, the wavevector
setting the modulation of the most singular charge fluctuations.
The  fermionic loop in Fig. 1(a,b) has the form
\beqa
\Lambda_{\alpha,\beta} (\Omega_l;\qvec,\omega_m)&=& CT\sum_n \sum_k 
\gamma_{\alpha,\beta}(\kvec)
G(\kvec,\varepsilon_n+\Omega_l)\nonumber \\
& \times & G(\kvec-\qvec,\varepsilon_n-\omega_m)
G(\kvec,\varepsilon_n),
\label{Ramvertex}
\eeqa
where $\gamma_{\alpha,\beta}(\kvec)\equiv \partial^2 
E_\kvec/\partial k_\alpha \partial k_\beta$ and $C$ is a constant 
determined by the coupling of the Raman vertex with the incoming and
outgoing photons and the coupling $g$ of the CM with the fermions. The choice
of the $\alpha$ and $\beta$ components depends on the
polarization of the incoming and outgoing photons \cite{reframan}.
A suitable choice of these polarizations corresponds to
specific projections of the $\gamma(\kvec)$ vertex on
cubic harmonics of the square lattice. To start our discussion
we chose the polarization corresponding to a Raman vertex
with $B_{1g}$ symmetry $\gamma_{B_{1g}}=\cos k_x-\cos k_y$
vanishing along the $(0,0)\to (\pm \pi,\pm\pi)$ directions. 
We are interested in the dominant contributions of the diagrams in Fig. 1(a,b),
which occur when the CM are around ${\bf q}_c$.
Therefore we set ${\bf q}={\bf q}_c$ in Eq. (\ref{Ramvertex}). In the
$k$ summation the largest contribution is obtained when
the three fermions are around the so-called hot spots, that is those regions
on the Fermi surface which can be connected by ${\bf q}_c$.  
For each given ${\bf q}_c$ in Eq. (\ref{Ramvertex}), to avoid cancellations, 
the sum over $k$ must encounter equivalent hot spots
where the vertex $\gamma_{B_{1g}}(k)$ does not change sign.
Since $\gamma_{B_{1g}}$ changes sign under the
$\pm x$ vs. $\pm y$ interchange, this condition is fulfilled
by stripes or eggbox CO fluctuations at not too small doping 
\cite{CDG,tranquada,yamada,STM},
where ${\bf q}_c\approx 2\pi(\pm 0.2,0),2\pi (0,\pm 0.2)$ [see Fig. 1(c)].
On the other hand, when for  the same ${\bf q}_c$ 
the  $B_{2g}$  vertex $\gamma_{B_{2g}}=\sin k_x \sin k_y$ 
is considered in Eq. (\ref{Ramvertex}),
the leading contribution from hot spots
vanishes since the equivalent ``available'' hot spots [see Fig. (1d)]
give contributions from regions where $\gamma_{B_{2g}}$ is opposite in sign.
With similar arguments one can show both in the superconducting \cite{venturini} and in the
normal phase, that spin CM at ${\bf q}_c=(\pm \pi, \pm \pi)$,
although having the similar hot spots as the CO ones, give rise to
non-vanishing vertices for the $A_{1g}$ symmetry only.
The same holds for spin CM with diagonal incommmensurate wavevectors
${\bf q}_c\approx (\pm (\pi-\delta), \pm (\pi-\delta))$, which is the case
of LSCO at low doping ($x \le 0.04$) \cite{wakimoto}.
On the other hand, at higher doping,
for small vertical and horizontal incommensuration
${\bf q}_c= (\pm (\pi-\delta), \pm \pi), (\pm \pi, \pm (\pi-\delta)) $, 
there is no complete cancellation in the ${B_{1g}}$  symmetry. This
small contribution could add to the CO contribution that we are going
to evaluate from the diagrams of Fig. 1(a,b), 
but cannot explain the anomalous Raman response in the deeply underdoped regime.

The resulting CM contribution to $B=B_{1g},B_{2g}$ Raman scattering is given by
\begin{eqnarray}
\Delta \chi_B ''&=&\Lambda_{B}^2 \int_0^{\infty}dz
\left[b(z-\omega/2)-b(z+\omega/2)\right] \nonumber \\
&\times& \frac{z_+z_- }{z_+^2-z_-^2}
\left[F(z_-)-F(z_+) \right] 
\end{eqnarray}
where $b(z)$ is the Bose function,
\beq
F(z)\equiv \frac{1}{z}\left[
\arctan\left(\frac{\omega_0}{z}\right) -
\arctan\left(\frac{m}{z}\right)
\right],
\eeq
and $z_\pm\equiv (z\pm \omega/2)(1+(z\pm \omega/2)^2/\omega_0^2)$.
Here $\omega_0\sim 100-500 \, cm^{-1}$ is an ultraviolet cutoff of the 
order of the frequency of the phonons most strongly coupled to the
electrons and driving the systems near the CO instability \cite{ACDG}.
The above expression of $\Delta \chi''$ is actually calculated
by considering CO collective modes with a semiphenomenological spectral density of the form
\beq
A(\omega,\Omega_\qvec)=\frac{\omega\left[1+\left(\frac{\omega}{\omega_0}\right)^2\right]}
{\omega^2\left[1+\left(\frac{\omega}{\omega_0}\right)^2\right]^2+\Omega_\qvec^2}
\label{spectrdens}
\eeq
In the limit of infinite $\omega_0$, one recovers the spectral
density of the above critical diffusive propagator $D(\qvec, \omega_m)$.

Experimentally \cite{tassini} in ${\rm La_{1.9}Sr_{0.1}CuO_4}$  
an anomalous Raman absorption is observed 
in the $B_{1g}$ symmetry, while the behavior of the
$B_{2g}$ spectra can be accounted for by QP only. Moreover the
anomaly in $B_{1g}$  softens upon lowering temperature and
tends to saturate suggesting a quasi-critical behavior. 
In Fig. 2(a) we report a comparison between our theoretical
calculations and experimental data. According to the neutron-scattering
experiments in this material \cite{tranquada,yamada}
we choose $\qvec_c$ along the (1,0) or (0,1)
directions, giving a non-vanishing vertex for the  $B_{1g}$ symmetry only.
We adjust at one temperature the 
overall intensity by choosing the vertex strength $\Lambda_{B_{1g}}$ of Eq. (\ref{Ramvertex})
and the ultraviolet cutoff $\omega_0$ to reproduce the 
lineshape. Then, keeping fixed these parameters
at all temperatures, we tune the mass $m(T)$ to reproduce all the curves.
The agreement is manifestly satisfactory, indicating that the
strong temperature dependence of the Raman absorption peaks 
is basically ruled by the temperature dependence of the 
low-energy scale $m(T)$ and by the temperature dependence
of the Bose factors.  The inset reports the
CM mass $m(x=0.10, T)$ needed to fit the experimental curves:
At  large to moderate temperatures $m(T)$ has a linear part,
which has an intercept on the $T$ axis at a temperature 
$T^*$ of the order of $70 \, K$. At lower temperatures
$m$ tends to saturate and the system crosses over
to a nearly ordered regime with a finite  $m(T)$. This behavior is
 clearly consistent with the behavior expected for the mass of
critical modes in the underdoped region: At large temperatures the
system is in the quantum critical regime with $m\propto (T-T^*)$, while 
below a crossover temperature $T^*$ [which for LSCO at $x=0.10$ is
indeed about $60-80 K$, see, e.g., Fig. (1a) of Ref.\onlinecite{ACDG}]
the CO transition is quenched by other effects.
\begin{figure}
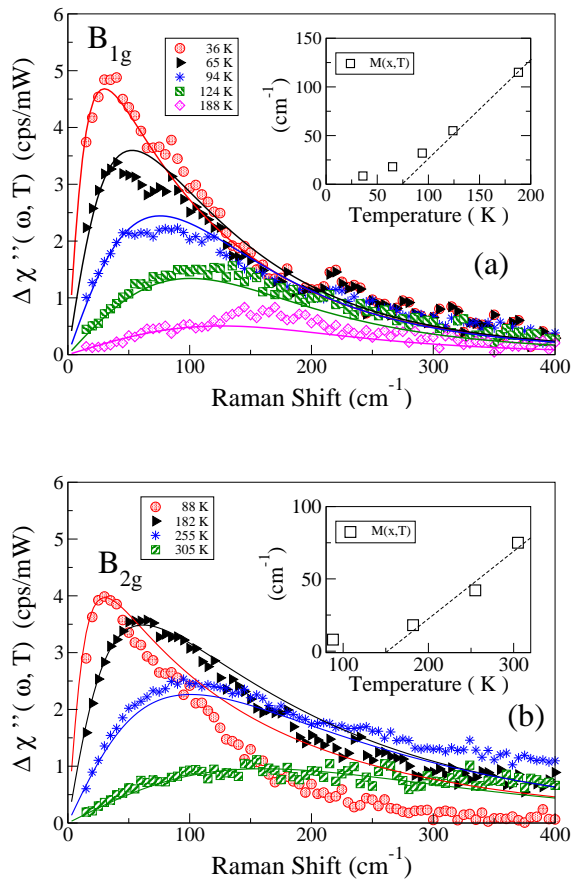

\includegraphics[angle=0,scale=0.3]{./Fig2a-raman.eps}%
\vspace{0.8 truecm}
\includegraphics[angle=0,scale=0.3]{./Fig2b-raman.eps}
\caption{(color online) (a) Comparison between experimental and calculated Raman 
spectra on ${\rm La_{1.90}Sr_{0.10}CuO_4}$. The intensity is chosen to reproduce the 
data with $\Lambda_{B_{1g}}^2=1.7$, the high-frequency cutoff is $\omega_0=250\, cm^{-1}$,
and the mass is reported in the inset.
(b) Same as (a) for ${\rm La_{1.98}Sr_{0.02}CuO_4}$ with  $\Lambda_{B_{2g}}^2=0.85$.
}
\label{fig.2}
%
\end{figure}

If the CM had wavevectors along the diagonal (1,1) and (1,-1)
directions and small modulus, the role of the $B_{1g}$ and $B_{2g}$ would be 
interchanged, with this latter displaying the anomalous Raman absorption.
Noticeably, experiments in ${\rm La_{1.98}Sr_{0.02}CuO_4}$, \cite{tassini},
where neutron scattering detect spin incommensurate (but likely also stripe)
order along the diagonal directions \cite{wakimoto}, do show that the anomalous Raman
absorption is present in the $B_{2g}$ symmetry and is absent in the
$B_{1g}$. Fig. 2(b) reports a comparison between experimental data
of Ref. \cite{tassini} and a calculated CM contribution for
this $B_{2g}$ case. In this case, consistenly with the expected
high value of $T^*$,  the temperature dependence of the mass
no longer displays a clear linear behavior and therefore the estrapolation 
of the high-temperature ``linear'' part to identify $T^*$ is not
well justified. Nevertheless, as a crude estimate from the last three points
we obtain $T^*$ of the order of 150 $K$, which is again consistent with the $T^*$
values reported in the literature (see, {\it e.g.}, Ref. \onlinecite{ACDG}). 
Notice also that the tail of the data at the lowest temperature ($T=88\, K$),
substantially lower than $T^*$, is not well reproduced by our calculations 
based on a diffusive form of the CM.

A similar behavior of the Raman spectra is observed in
${\rm (Y_{1.97}Ca_{0.3})Ba_2CuO_{6.05}}$ sample
\cite{hacklSNS04} at filling $p=0.03$, with a temperature-dependent peak in 
the $B_{2g}$ symmetry only. Again it is quite natural to attribute this
peak to stripe fluctuations along the $(1,\pm 1)$  directions and long wavelength
$[{\bf q}_c \approx (p,p)]$. Fig. 3 reports the comparison between experiments and
our theoretical analysis. Again a value of $T^*\sim 120\, K$ may be estrapolated from
the high temperature mass behavior.

The data for YCBCO also display a remarkable change in the lineshape at the lowest
temperatures (more pronounced than in ${\rm La_{1.98}Sr_{0.02}CuO_4}$ at
$T=88\, K$. In particular the peak at $T=11.0\, K$
is much narrower and decreases more rapidly at high frequencies. This change
in the lineshape naturally marks a change in the nature of the charge
fluctuations from the the high-temperature diffusive behavior in the
quantum critical regime to a sharper mode behavior
typical of the low-temperature underdoped regime. In this case
the diffusive form of the propagator fails in reproducing the 
lineshape, while a good description may be obtained by 
a factorized form, according to
previous treatments of the nearly ordered underdoped regime
\cite{KAMPF}. Therefore  we also carried out the calculations using
\begin{equation}\label{chi}
{\tilde D}({\bf q}, \omega_n)=W(i\omega_n)J_x({\bf q})J_y({\bf q}),
\end{equation}
where
$
J_{x,y}({\bf q})={\cal N}
\gamma\left[\gamma^2+(q_{x,y}-q_{cx,y})^2 \right]^{-1}.
$
Here ${\cal N}$ is a normalization factor and
the inverse correlation length is $\gamma$. The (real-frequency)-dependent part 
$W(\omega)=\int d\nu {\tilde A}(\nu) 2\nu/(\omega^2-\nu^2)$
with ${\tilde A}(\omega)\sim \Gamma/[(\omega-\omega_{CM})^2 +\Gamma^2]$
is a normalized lorentzian  centered around 
the typical CM energy $\omega_{CM}$ with halfwidth $\Gamma$.
$\Delta \chi_B ''$ now reads
$\Delta \chi_B ''=B(T)f(\omega/\Gamma,T/\Gamma,\omega_{CM}/\Gamma)$
with $B(T) \propto {\tilde g}(T)^4/[\gamma(T)^2\Gamma(T)]$, $\tilde g$
being the coupling between the CM and the QP \cite{notacoupling}.
As shown by the dashed curves in Fig. 3, the
factorized ``propagator'' provides a good description at the lowest temperature,
while in the intermediate-temperature regime ($T=86.2\, K$, $T=127.3\, K$)
the diffusive propagator becomes progressively a better description of the
charge fluctuations. We also notice that the mass $m(T)$
of the quantum-critical diffusive propagator smoothly connects with the
energy of the low-temperature modes: $m(T) \sim \omega_{CM}$. 

As far as LSCO samples are concerned, the diffusive form of the modes
seems to persist down to the lowest temperatures in the available published data.
\begin{figure}
\includegraphics[angle=0,scale=0.3]{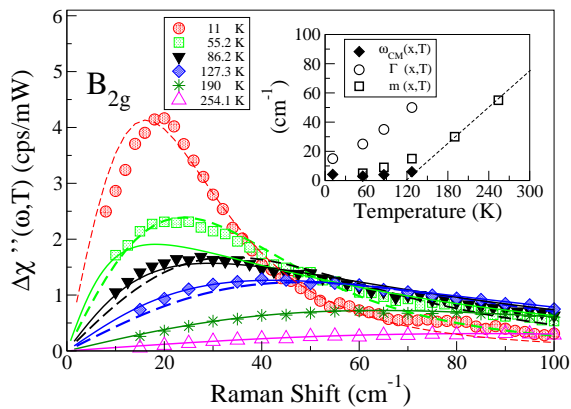}
\caption{(Color online) Comparison between experimental and
calculated Raman spectra on ${\rm (Y_{1.97}Ca_{0.3})Ba_2CuO_{6.05}}$
with diffusive propagators [Eq. (\ref{diffusive})] (solid lines).
The intensity is chosen to fit the 
data with $\Lambda_{B_{2g}}^2=0.41$, the high-frequency cutoff is $\omega_0=100\, cm^{-1}$,
and the mass is reported in the inset. 
The dashed lines are obtained with factorized ``propagators'' 
[Eq. (\ref{chi})] with overall intensity 
$B(T)=28,\,\,\,27, \,\,\,
19,\,\,\, 12$ (cps/mW) for $T=11\, K, \,\,\, 55.2\,K,\,\,\,86.2\, K,\,\,\,127.3\,K$.
$\Gamma(T)$ and $\omega_{CM}$ are also reported in the inset.
}
\label{fig.3}
%
\end{figure}
We notice in passing that the two values for $T^*$ obtained
at $x=0.02$ and $x=0.10$ linearly estrapolate to $T^*=0$
at $x=x_c\approx 0.17$, 
consistent with the theoretical position of the CO quantum critical point 
\cite{CDG,ACDG} and not inconsistent with most of the experiments on
$T^*$ in LSCO \cite{reviewQCP}. 

In order to emerge clearly from the
other electronic excitations, the CM's need to have rather small masses
(from the insets of Figs. 2 and 3 a few tens of $cm^{-1}$).
This is the case in underdoped systems 
where the finite and sizable values of $T^*$ carry to finite temperature
the small values of the mass, which should  vanish at $T_{CO}(x)\sim T^*$
if  a genuine long-range order would occur. Around optimal doping, where 
$T_{CO}(x_c)=0$, anomalous peaks are hardly visible
because superconductivity intervenes at higher temperatures
before the mass, being proportional to $T$, becomes  sufficiently small.
Similarly we argue that the observation of CM peaks in Bi-2212
systems is unlikely. In these materials, STM experiments suggest that
the CO coherence lenght is smaller than 5-6 lattice spacings \cite{STM}, indicating
large mass values, in contrast to underdoped LSCO materials, where 
according to neutron scattering experiments the stripe
correlations may extend over tens of lattice spacings \cite{tranquada}.

In summary we calculated the CO-CM contribution to the Raman response function.
The idea that charge modes are important in the low-energy Raman spectra is also
supported by recent experiments in ladder compounds \cite{gozar}.
Here we showed that the CM contribution quantitatively accounts for
the observed anomalous peaks in Raman spectra of LSCO and YCBCO. In this explanation
a crucial role is played by the symmetry of the Raman vertices: 
We demonstrate the stringent connection between the symmetry properties of the
Raman spectra and the doping, which implies a 
dependence of the CM on {\it finite} wavevectors specific of 
stripe or eggbox fluctuations as inferred from neutron scattering.
This rules out for the interpretation of these Raman experiments
local ({\it i.e.}, at all momenta) excitations like
polarons, disorder-localized single particles or resonating-valence-bond
singlets. Also spin CM do not
comply with these symmetry requirements over the whole doping range examined above.
For the same reason other excitations peaked 
at $\qvec=0$ (like superconducting pair fluctuations
or time-reversal-breaking current fluctuations) are not appropriate. 
Although we cannot {\it a priori}
exclude that the proposed CO collective excitations are concomitant with these other more
elusive forms of criticality, at the moment the anomalous Raman absorption
can only be interpreted with CM peaked at finite momenta.

We acknowledge interesting discussions with C. Castellani, T. Devereaux, and J. Lorenzana.
We particularly thank R. Hackl and L. Tassini for discussions and for providing us the
Raman data before publication. We acknowledge financial support form the MIUR-COFIN2003
n. 2003020230\_006. C.D.C. thanks the Walther Meissner Institute in Garching and the
Humboldt Fundation for hospitality and support. 



\begin{thebibliography}{99}
\bibitem{notacoupling} The intensity of the spectra $B(T)$ has a temperature 
dependence through $\gamma(T)$, $\Gamma(T)$, and $\tilde g(T)$  This latter
in the ``factorized-propagator'' scheme  is proportional to the 
(local) order parameter \cite{KAMPF}. These dependencies saturate at low T and
lead to an increasing and saturating $B(T)$ for $T$ much smaller than $T^*$.
\end{thebibliography}
\end{document}